\documentclass[12pt,showpacs,preprintnumbers]{revtex4}
 \usepackage{amsmath}
 \usepackage{amssymb}
 \usepackage{epsfig}
 \usepackage{graphicx}
 
\begin{document}

\preprint{BIHEP-TH-2003-27}

\title{Rare Decays  $B^0 \to D_s^{(*)+} D_s^{(*)-}$ and
$B_s^0 \to D^{(*)+} D^{(*)-}$ in Perturbative QCD Approach}
\author{Ying Li$^{a,b,c}$\thanks{e-mail: liying@mail.ihep.ac.cn},
 Cai-Dian L\"u$^{a,b}$\thanks{e-mail:  lucd@ihep.ac.cn},
 and Zhen-Jun Xiao$^{a,c}$\thanks{e-mail: zjxiao@email.njnu.edu.cn} \\
{\it \small $a$ CCAST (World Laboratory), P.O. Box 8730,
Beijing 100080, China}\\
 {\it \small $b$   Institute of High Energy Physics,
CAS, P.O.Box 918(4)
  Beijing 100039, China}\\
{\it \small $c$ Physics Department, NanJing Normal University,
JiangSu 210097, China}}

\begin{abstract}
In decay modes $B^0 \to D_s^{(*)+} D_s^{(*)-}$ and $B_s^0 \to
D^{(*)+} D^{(*)-}$, none of quarks in final states is the same as
one of $B(B_s)$ meson. They can occur only via annihilation
diagrams in the Standard Model. In the heavy quark limit, we try
to calculate the branching ratios of these decays in perturbative
QCD approach without considering the soft final state interaction.
We found branching ratios of $B^0 \to D_s^{(*)+} D_s^{(*)-}$ are
at the order of $ 10^{-5}$, and branching ratios of $B_s^0 \to
D^{(*)+} D^{(*)-}$ are of $10^{-3}$. Those decay modes will be
measured in $B$ factories and LHC-b experiments.
\end{abstract}


\maketitle

\section{Introduction}
As an important way in testing the Standard Model and searching
for new physics, rare $B$ decays become important in particle
physics. Although some of them have been measured by $B$
factories, many of them are still under study from both
experimental and theoretical sides. In theoretical side, the
factorization approach has been accepted because it can explain
many decay branching ratios successfully. Recently many efforts
have been made to explain the reason why the factorization
approach has worked well. One of them is perturbative QCD
approach (PQCD) \cite{pqcd}, in which we can calculate the
annihilation diagrams as well as the factorizable and
non-factorizable diagrams. It has been applied to exclusive $B$
meson decays, such as $B \to \pi\pi(\rho)$ \cite{luy}, $B \to
K\pi$ \cite{kls}, $B \to D_s^{(*)}K$ \cite{lu:bdsk} and some other
channels \cite{pqcd2,power,dpi}.

\begin{figure}[htbp]
  \epsfig{file=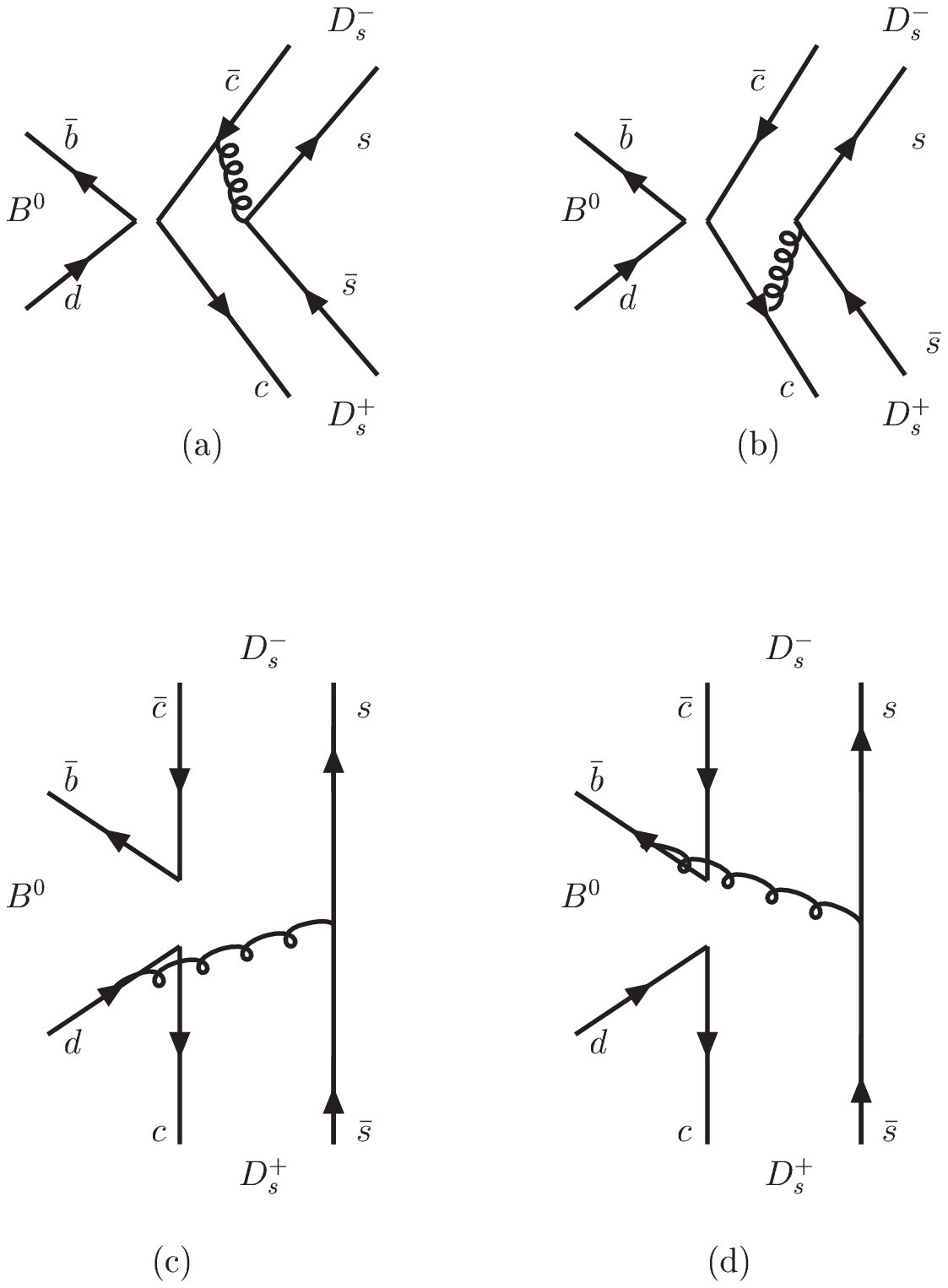,bbllx=5cm,bblly=10cm,bburx=10cm,bbury=26cm,%
 width=3cm,angle=0}
 \caption{Diagrams for $B^0 \to D_s^+D_s^-$ decay. The factorizable
 diagrams (a) and (b) contribute to $F_a$, and the nonfactorizable diagrams (c) and
 (d) do to $M_a$.}
 \label{fig:diagrams1}
 \end{figure}

Recently,  J. O. Eeg $et\  al$. computed the $B^0 \to D_s^+ D_s^-$
and $B_s^0 \to D^+ D^-$ decays using heavy-light Chiral quark
model which is a non-perturbative approach \cite{eeg}. As shown in
Fig.\ref{fig:diagrams1}, the four quarks in final states $D_s^+$
and $D_s^-$ are different from the ones in the $B$ meson, and
there is no spectator quark. So this decay is a pure annihilation
type decay. In the factorization approach, this decay is described
as $b$ and $\bar d$ in $B$ meson annihilation into vacuum and
$D_s^+$, $D_s^-$ being produced from vacuum afterwards. If we
calculate this decay in factorization approach \cite{bsw}, we need
the $D_s^+ \to D_s^-$ form factor at very large momentum transfer
${\cal O} (m_B^2)$, but it is zero due to vector current conservation. So it is difficult to
calculate this decay model in factorization approach.
                    In the so called QCD factorization approach
                    \cite{bbns}, the annihilation contribution is
                    plagued by the endpoint singularity. Thus it
                    is only parameterized as a free parameter for this
                    kind of contribution. On the other hand, by including the transverse
                    momentum of the partons, the
                    PQCD approach is free of such singularities.
                    Furthermore, the Sudakov factor induced by the
                    inclusion of transverse momentum helps the
                    convergence of factorization.

PQCD approach has been recently applied to $B$ meson decays with
one charmed meson in the final states \cite{lu:bdsk,dpi}.
In the typical $B\to D\pi$ decays, the momentum of the final state
meson is approximately $\frac{1}{2} m_B (1- r^2 )$, with
$r=m_D^2/m_B^2$. This is still  large enough to make a
hard intermediate gluon   in the hard part calculation. Therefore
the  predicted
results  in PQCD agree well with the experimental data. The PQCD calculation of  $B^0 \to
D_s^{(*)+} D_s^{(*)-}$ and $B_s^0 \to D^{(*)+} D^{(*)-}$ decays
with two charmed mesons in final states may  be  questionable,
since the momentum of final state meson    here  is relatively
smaller than that of $B\to D\pi$ case. However, after calculation
we find that the momentum of final state $D_{(s)}$ meson is
$\frac{1}{2}m_B\sqrt{1- 4 r^2} \simeq \frac{1}{2}
m_B (1-2 r^2)$, which is only a little smaller than that of $B\to D\pi $ case.
 For example, the $W$
boson exchange causes $\bar{b}d \to \bar{c}c $, and the $\bar{s}s$
quarks are produced from a gluon. This gluon attaches to any one
of the quarks participating in the $W$ boson exchange.
In the heavy quark limit, we apply the hierarchy approximation,
which is adopted by ref.\cite{power,dpi}, $\Lambda_{QCD} \ll m_D
\ll m_B$. In this limit, the D meson momentum is nearly $m_B/2$.
According to the distribution amplitude used in ref.\cite{dpi},
the light quark in D meson carrying nearly $40\%$ of the D meson
momentum. It
is still a collinear quark with 1 GeV energy, like that
in $B\to D\pi$, $B\to \pi\pi$ decays.
Therefore the gluon connecting them is a
hard gluon, so we can perturbatively treat the process where the
four-quark operator exchanges a hard gluon with $s \bar s$ quark
pair.

The framework of PQCD and analytic formulas for the decay
amplitudes will be shown in the next section. In section
\ref{sc:neval}, we give the numerical results and discussion.
Finally, we conclude this study in section \ref{sc:concl}.

\section{Framework}

The factorization theorem allows us to separate the decay
amplitude into soft($\Phi$), hard($H$), and harder ($C$) dynamics
characterized by different scales \cite{luy,kls}. It is expressed as
\begin{equation}
 \mbox{Amplitude}
\sim \int\!\! d^4k_1 d^4k_2 d^4k_3\ \mathrm{Tr} \bigl[ C(t)
\Phi_B(k_1) \Phi_{D}(k_2) \Phi_D(k_3) H(k_1,k_2,k_3, t) \bigr],
\label{eq:convolution1}
\end{equation}
where $k_i$'s are momenta of light quarks included in each meson,
and $\mbox{Tr}$ is the trace over Dirac and color indices.
 The soft dynamic is factorized into
 the meson wave function $\Phi_M$, which describes hadronization of
 the quark and anti-quark pair into the meson $M$.
The
harder dynamic involves the four quark operators described by the
Wilson coefficient $C(t)$.    It results from the radiative
corrections to the four quark operators at short distance.
 $H$ describes the four
quark operator and the quark pair from the sea connected by a hard
gluon whose scale is at the order of $M_B$, so the hard part $H$ can be
perturbatively calculated. The hard and harder dynamics together make
an effective six quark interaction. The $H$ depends on the specific
process, while $\Phi_M$ is independent of any processes.
Therefore we may determine $\Phi_M$ by other well measured
channels to make prediction here.

We consider the $B$ meson at rest for simplicity. It is convenient
to use light-cone coordinates $(p^+,p^-,{\bf p}_T)$, which is
defined as:
\begin{equation}
       p^+ =\frac{p^0+p^3}{\sqrt 2} , \ \ p^- =\frac{p^0-p^3}{\sqrt
       2},\ \ {\bf p}_T =(p^1,p^2).
       \label{lightcone}
\end{equation}
Thus, expanding up to the order of $r^2$, we can take  the $B$ meson and two $D^{(*)}$
meson momenta
as:
\begin{equation}
       P_1 = \frac{M_B}{\sqrt{2}} (1,1,{\bf 0}_T), P_2 =
       \frac{M_B}{\sqrt{2}} (1-r^2,r^2,{\bf 0}_T), P_3 =
       \frac{M_B}{\sqrt{2}} (r^2,1-r^2,{\bf 0}_T) , \label{eq:momentun1}
\end{equation}
where $r= M_{D^{(*)}}/M_B$. Putting the light (anti-)quark momenta
in $B$, $D^+$, $D^-$ mesons as $k_1$, $k_2$, and $k_3$,
respectively, we can choose
\begin{equation}
k_1 = (x_1P_1^+,0,{\bf k}_{1T}),\ \  k_2 = (x_2 P_2^+,0,{\bf
k}_{2T}),\ \
 k_3 = (0, x_3 P_3^-,{\bf k}_{3T}). \label{eq:momentun2}
\end{equation}
  Unlike QCD factorization approach, we do not neglect the transverse
 momentum $k_T$  in the above expressions, by which to avoid
 the endpoint singularity.

If the decay involves  one or two vector mesons in  final states, the
longitudinal polarization vectors $\epsilon_2^L$, $\epsilon_3^L$
up to the order of $r^2$
are  given by
\begin{equation}
\epsilon^L_2= \frac{M_B}{M_{D^*}\sqrt{2}} (1-r^2,-r^2,{\bf 0}_T),\
 \epsilon^L_3 =\frac{M_B}{M_{D^*}\sqrt{2}} (-r^2,1-r^2,{\bf 0}_T),
\end{equation}
and transverse polarization vectors $\epsilon_2^T$, $\epsilon_3^T$
are
\begin{equation}
 \epsilon^T_2=(0,0,{\bf 1}_T),\ \
 \epsilon^T_3 =(0,0,{\bf 1}_T). \label{eq:pola}
\end{equation}
Then, integration over $k_1^-$, $k_2^-$, and $k_3^+$ in
eq.(\ref{eq:convolution1}) leads to:
\begin{multline}
 \mbox{Amplitude}
\sim \int\!\! d x_1 d x_2 d x_3
b_1 d b_1 b_2 d b_2 b_3 d b_3 \\
\mathrm{Tr} \bigl[ C(t) \Phi_B(x_1,b_1) \Phi_{D}(x_2,b_2)
\Phi_D(x_3, b_3) H(x_i, b_i,t) S_t(x_i)\, e^{-S(t)} \bigr],
\label{eq:convolution2}
\end{multline}
where $b_i$ is the conjugate space coordinate of the transverse momentum $k_{iT}$, and $t$
is the largest energy scale in $H$, as a function in terms of
$x_i$ and $b_i$. The last term, $e^{-S(t)}$, contains two kinds of
logarithms. One of the large logarithms is due to the
renormalization of ultra-violet divergence $\ln tb$, the other is
double logarithm $\ln^2 b$ from the overlap of collinear and soft
gluon corrections. This Sudakov form factor suppresses the soft
dynamics effectively \cite{soft}, so it makes a perturbative
calculation of hard part $H$ applicable at the intermediate scale.

As a heavy meson, the $B$ meson wave function is  not well
defined, so is $D$ meson. In heavy quark limit, we may use only
one independent distribution amplitude for each of them.
\begin{equation}
 \Phi_{B}(x,b) = \frac{i}{\sqrt{6}}
\left[ \not \! P  + M_B  \right] \gamma_5\phi_B(x,b),
 \end{equation}
 \begin{equation}
 \Phi_{D}(x,b) = \frac{i}{\sqrt{6}}\gamma_5
\left[ \not \! P  +  M_D \right] \phi_D(x,b).
\end{equation}
For the vector $D^*$ meson, it is expressed as:
\begin{equation}
 \Phi_{D^*}(x,b) = \frac{i}{\sqrt{6}}
 \not \! \epsilon\left[ \not \! P+ M_{D^*} \right ]
\phi_{D^*}(x,b).
\end{equation}

The hard part $H$, which is channel dependent, can be calculated
perturbatively. We show the calculated formulas below for
different channels.

\subsection{$B^0 \to D_s^+ D_s^-$, $B^0_s \to D^+ D^-$ decays}{\label{s1}}

In the decay $B^0 \to D_s^+ D_s^-$, the effective Hamiltonian at
scale lower than $M_W$ is \cite{Buchalla:1996vs}:
\begin{gather}
 H_\mathrm{eff} = \frac{G_F}{\sqrt{2}} V_{cb}^*V_{cd} \left[
C_1(\mu) O_1(\mu) + C_2(\mu) O_2(\mu) \right], \label{eff1}\\
  O_1 = (\bar{d}b)_{V-A} (\bar{c}c)_{V-A}, \quad
 O_2 = (\bar{c}b)_{V-A} (\bar{d}c)_{V-A}.\label{op1}
\end{gather}
In above functions, $C_{1,2}(\mu)$ are Wilson coefficients at
renormalization scale $\mu$. And summing over  $\mathrm{SU}(3)_c$
color's index $\alpha$, $\sum_\alpha \bar{q}_\alpha
\gamma^\nu(1-\gamma_5)q'_\alpha$, are abbreviated to
$(\bar{q}q')_{V-A}$. Penguin operators may also have contribution,
but they usually have smaller Wilson coefficients. Here we neglect these
diagrams. The lowest order diagrams for the hard part $H$
calculation,   are drawn in Fig.\ref{fig:diagrams1} according to
this effective Hamiltonian. Just as what we said above, there are
only annihilation diagrams.

For the decay $B^0_s \to D^+ D^-$, the effective Hamiltonian at
scale lower than $M_W$ is \cite{Buchalla:1996vs}:
\begin{gather}
 H_\mathrm{eff} = \frac{G_F}{\sqrt{2}} V_{cb}^*V_{cs} \left[
C_1(\mu) O_1'(\mu) + C_2(\mu) O_2'(\mu) \right], \label{eff1b}\\
  O_1' = (\bar{s}b)_{V-A} (\bar{c}c)_{V-A}, \quad
 O_2' = (\bar{c}b)_{V-A} (\bar{s}c)_{V-A}.\label{op1b}
\end{gather}
Comparing with  Eqs.(\ref{eff1},\ref{op1}), the only changes in Eqs.(\ref{eff1b},\ref{op1b})
 are the replacements of the CKM factor $V_{cd} \to V_{cs}$ and the quark $d \to s$.
As we will see later the branching ratio of $B^0_s \to D^+ D^-$
will be much larger than that of $B^0 \to D_s^+ D_s^-$ decay,
because of this larger CKM factor $V_{cs}$.
 The lowest order
diagrams for the hard part $H$ calculation,   are then similar to
$B^0 \to D_s^+ D_s^-$ decay  in Fig.\ref{fig:diagrams1}  only
replacing the $d$ quark by $s$ quark.

In decay $B^0 \to D_s^+ D_s^-$, we get the following analytic
formulas by calculating the hard part $H$ at first order in
$\alpha_s$.
The factorizable annihilation diagrams in Fig.\ref{fig:diagrams1}a
and b cancels each other,
 which is a result of conservation of vector current and parity
 invariance.

With the meson wave functions, the decay amplitude for
 the nonfactorizable annihilation diagrams in
Fig.\ref{fig:diagrams1}(c) and (d) results in
\begin{multline}
M_{a}  =  \frac{1}{\sqrt{6}} 64\pi C_F M_B^2 \int_0^1 \!\! dx_1
dx_2 dx_3
 \int_0^\infty \!\! b_1 db_1\, b_2 db_2\
\phi_B(x_1,b_1) \phi_{D_s}(x_2,b_2)\phi_{D_s}(x_3,b_2) \\
\times \Bigl[ -x_3 E_{m}(t_{m}^1) h_a^{(1)}(x_1, x_2,x_3,b_1,b_2)
+x_2 E_{m}(t_{m}^2) h_a^{(2)}(x_1, x_2,x_3,b_1,b_2) \Bigr],
\label{eq:Ma1}
\end{multline}
where $C_F = 4/3$ is the group factor of $\mathrm{SU}(3)_c$ gauge
group.
 The function  $E_m$ is defined as
 \begin{gather}
  E_{m}(t) = C_2(t) \alpha_s(t)\, e^{-S_B(t)-S_D(t)-S_D(t)},
 \end{gather}
  where
  $S_B$, $S_D$ result from summing  double logarithms caused by
 infrared gluon corrections and single logarithms  due to the
 renormalization of ultra-violet divergence \cite{pqcd2}.

 The functions $h_a^{(1)}$ and $h_a^{(2)}$
  are the Fourier transformation
 of virtual quark and gluon propagators.     They are
 defined by
 \begin{align}
 &h^{(j)}_a(x_1,x_2,x_3,b_1,b_2) = \nonumber \\
 & \biggl\{ \frac{\pi i}{2}
 \mathrm{H}_0^{(1)}(M_B\sqrt{x_2x_3(1-2r^2)}\, b_1)
  \mathrm{J}_0(M_B\sqrt{x_2x_3(1-2r^2)}\, b_2) \theta(b_1-b_2)
 \nonumber \\
 & \qquad\qquad\qquad\qquad + (b_1 \leftrightarrow b_2) \biggr\}
  \times\left(
 \begin{matrix}
  \mathrm{K}_0(M_B F_{(j)} b_1), & \text{for}\quad F^2_{(j)}>0 \\
  \frac{\pi i}{2} \mathrm{H}_0^{(1)}(M_B\sqrt{|F^2_{(j)}|}\ b_1), &
  \text{for}\quad F^2_{(j)}<0
 \end{matrix}\right),
 \label{eq:propagator2}
 \end{align}
 where $\mathrm{H}_0^{(1)}(z) = \mathrm{J}_0(z) + i\,
 \mathrm{Y}_0(z)$, and $F_{(j)}$s are defined by
 \begin{equation}
  F^2_{(1)} = x_1x_3(1-r^2)-x_2x_3(1-2r^2),
 \end{equation}
 \begin{equation}
  F^2_{(2)} =x_1+x_2x_3+(1-r^2)(x_2+x_3-x_1x_3-2x_2x_3).
 \end{equation}
  The hard scale
 $t$'s in the amplitudes are taken as the largest energy scale in
 the $H$ to kill the large logarithmic radiative corrections:
 \begin{gather}
  t_{m}^j = \mathrm{max}(M_B \sqrt{|F^2_{(j)}|},
 M_B \sqrt{(1-2r^2)x_2x_3 }, 1/b_1,1/b_2).
 \end{gather}

Applying the power counting rule established in ref.\cite{power,dpi},
we keep only the leading order contribution of $r$ expansion in the numerator of
the above equation (\ref{eq:propagator2}).
 The hierarchy
relation $\Lambda_{QCD} \ll m_{D^{(*)}} \ll m_B$  is assumed.
 The $r^2$ terms are kept in the denominators   of
 (\ref{eq:propagator2}), since it may sometimes affect  the imaginary
 part heavily.   It is easy to see that the momentum carried by
 the intermediate gluon is $M_B\sqrt{x_2x_3(1-2r^2)}$ which is
 only suppressed by a factor of $\sqrt{(1-2r^2)}$, comparing with
 that of
 the charmless  $B\to \pi\pi$ decay \cite{luy}. In the heavy quark limit, $r\to 0$,
  the momentum of the gluon is the same for the two kinds of
  decays.   The formulas derived here support the argument at the
  introduction that perturbative calculation is still applicable
  to the $B\to D D$ decays.

 The decay width $\Gamma$ for $B^0 \to D_s^+
D_s^-$ decay is then given by
\begin{equation}
 \Gamma(B^0 \to D_s^+ D_s^-) = \frac{G_F^2 M_B^3}{128\pi} (1-2r^2)
\bigl|V_{cb}^*V_{cd}M_{a}\bigr|^2. \label{eq:neut_width1}
\end{equation}
Similar to decay $B^0 \to D_s^+ D_s^-$, the width for $ B^0_s \to
D^+ D^-$ is
\begin{equation}
 \Gamma(B^0_s \to D^+ D^-) = \frac{G_F^2 M_B^3}{128\pi} (1-2r^2) \bigl|V_{cb}^*V_{cs}
M_{a}\bigr|^2. \label{eq:neut_width2}
\end{equation}
One need only replace the $D_s$ wave function $\phi_{D_s}$ by $D$
meson. Enhanced by the  CKM factor $|V_{cb}^*V_{cs}|^2$, the $B_s^0$
decay width will be  larger than that of $B^0$ decay.

\subsection{$B^0 \to D_s^{*+} D_s^-$, $D_s^{+} D_s^{*-}$ and $B_s^0 \to D^{*+} D^{-}$, $D^{+} D^{*-}$ decays}

For final states with one pseudo-scalar and one vector mesons, only
the longitudinal polarization of vector meson contribute. The
decay amplitude takes the same form as the amplitude of $B$ to two
pseudo-scalar mesons (\ref{eq:Ma1}). For decay $B^0
\to D_s^{*+} D_s^-$, one need only replace one of the $D_s$ meson
distribution amplitude by $D_s^*$ one. The width of $B^0 \to
D_s^{+} D_s^{*-}$ must have the same width as $B^0 \to D_s^{*+}
D_s^{-}$. The contributions of Fig1.(a) and (b) can  not be
cancelled by each other because of the difference of $f_{D_s}$ and
$f_{D_s^*}$, but it is still negligible.

Accordingly, the $B_s^0 \to D^{*+} D^{-}$, $D^{+} D^{*-}$ decay
amplitudes also take the same form as $B_s^0 \to D^{+} D^{-}$.

\subsection{$B^0 \to D_s^{*+} D_s^{*-}$ and $B_s^0 \to D^{*+} D^{*-}$ decays}

There are contributions not only  from the longitudinal
polarization but also from two transverse polarizations in $B\to
VV$ decays, where $V$ denotes the vector meson. Therefore the
decays $B^0 \to D_s^{*+} D_s^{*-}$ and $B_s^0 \to D^{*+} D^{*-}$
are more complicated than $B\to PP$ or $B\to PV$.

In the covariant form, the decay amplitudes of non-factorizable
annihilation diagrams are
\begin{multline}
M_a'=\frac{1}{\sqrt{2N_c}}128\pi r^2  C_F \int_0^1 \!\! dx_1 dx_2
dx_3
 \int_0^\infty \!\! b_1 db_1\, b_2 db_2\
\phi_B(x_1,b_1) \phi_{D_s^*}(x_2,b_2) \phi_{D_s^*}(x_3,b_2)\\
\times\biggl\{(\epsilon_2\cdot\epsilon_3) (p_3 \cdot p_2)(x_2+x_3)
\Bigl[E_{m}(t_{m}^1) h_a^{(1)}(x_1,
x_2,x_3,b_1,b_2) - E_{m}(t_{m}^2) h_a^{(2)}(x_1, x_2,x_3,b_1,b_2) \Bigr]\\
+ (\epsilon_2 \cdot p_3)(\epsilon_3 \cdot p_2) (x_2-x_3) \Bigl[
E_{m}(t_{m}^1) h_a^{(1)}(x_1, x_2,x_3,b_1,b_2) + E_{m}(t_{m}^2)
h_a^{(2)}(x_1,
x_2,x_3,b_1,b_2) \Bigr]\\
-i \varepsilon^{\mu\nu\rho\sigma}\epsilon_{2\mu}\epsilon_{3\nu}
p_{2\rho} p_{3\sigma} (x_2-x_3) \Bigl[ E_{m}(t_{m}^1)
h_a^{(1)}(x_1, x_2,x_3,b_1,b_2)-E_{m}(t_{m}^2) h_a^{(2)}(x_1,
x_2,x_3,b_1,b_2)
  \Bigr]\biggr\}, \label{eq:Ma32}
\end{multline}
with the convention $tr(\gamma_5 \not \! a \not \! b \not \! c
\not \! d)=-4i\epsilon^{\alpha\beta\gamma\rho}a_\alpha b_\beta c_
\gamma d_\rho$ and $\epsilon^{0123}=1$. Just as
Section (\ref{s1}), the contributions of diagrams(a) and (b)
cancel each other.

 If we set the $\epsilon_i$ to be longitudinal polarization only, the above
 formula goes back to the eq.(\ref{eq:Ma1}).
  From the above functions, we can also see that the contributions of transverse
polarizations are proportional to factors of $r^2$, which are
suppressed comparing with longitudinal ones.
 In our
calculation, we set $m_c\approx m_{D_{(s)}^{(*)}}$, just because
$m_{D_{(s)}^{(*)}}-m_c\sim\bar\Lambda$. And
$\Lambda/m_{D_{(s)}^{(*)}} \to 0$ in the heavy quark limit.

\section{Numerical Results}\label{sc:neval}

For $B$ meson, we use the same wave functions as other
 charmless $B$ decays \cite{luy,kls}, which is chosen as
\begin{equation}
\phi_B(x,b) = N_B x^2(1-x)^2 \exp \left[ -\frac{M_B^2\ x^2}{2
\omega_b^2} -\frac{1}{2} (\omega_b b)^2 \right].\label{waveb}
\end{equation}
The parameters  $\omega_b=0.4\mbox{ GeV}$, and
$N_B=91.745\mbox{GeV}$ which is the normalization constant using
$f_B=190 \mbox{MeV}$, are constrained by charmless B decays
\cite{luy,kls}. For $B_s$ meson, we use the same wave function
according to SU(3) symmetry. That is $\omega_b=0.4\mbox{ GeV}$,
but $N_{B_s}=119.4\mbox{GeV}$, using $f_{B_s}=236 \mbox{MeV}$.

For $D_{(s)}^{(*)}$, the distribution amplitude  is taken as \cite{lu:bdsk,dpi}
\begin{equation}
\phi_{D_{(s)}^{(*)}}(x,b) = \frac{3}{\sqrt{2 N_c}}
f_{D_{(s)}^{(*)}} x(1-x)\{ 1 + a_{D_{(s)}^{(*)}} (1 -2x)
\}.\label{waved}
\end{equation}
Since the heavy $D_{(s)}^{(*)}$ wave function is less constrained,
we use $a_{D^{(*)}}=0.6\sim 0.8\mbox{ GeV}$ and $a_{D_s^{(*)}}=0.2\sim 0.4 \mbox{
GeV}$ to explore the sensitivity of parameters. Other parameters,
such as meson mass, decay constants, the CKM matrix elements and
the lifetime of $B_{(s)}$ meson \cite{pdg} are given in
Table~\ref{para}.

 \begin{table}[htbp]
\caption{Parameters we used in numerical calculation.} \label{para}
\begin{center}
\begin{tabular}[t]{r|cc}
 \hline     \hline
   \     &$m_{B^0}=5.28\mbox{GeV}$ & $m_{B_s^0}=5.37 \mbox{GeV} $ \\
 Mass    &$m_{D^{\pm}}=1.87 \mbox{GeV}$  &$m_{D^{*\pm}}=2.01\mbox{GeV}$  \\
   \     &$m_{D_s^{\pm}}=1.97 \mbox{GeV}$ & $m_{D_s^{*\pm}}=2.11 \mbox{GeV}$  \\
 \hline
 \hline
 Decay    & $f_{D^{\pm}}=240 \mbox{ MeV}$ &$f_{D^{*\pm}}=230 \mbox{MeV}$  \\
 Constants& $f_{D_s^{\pm}}=241 \mbox{MeV}$ &$f_{D_s^{*\pm}} = 211 \mbox{ MeV}$  \\
 \hline
 \hline
 CKM   & $|V_{cd}|= 0.224$ & $|V_{cb}|= 0.042$ \ \\
 & $|V_{cs}|= 0.974$ \ \\
 \hline
  \hline
Lifetime  & $\tau_{B^0}=1.54\times 10^{-12}\mbox{ s}$
\         & $\tau_{B_s^0}=1.46\times 10^{-12}\mbox{s}$ \ \\
 \hline
\end{tabular}
\end{center}
\end{table}

The calculated branching ratios in PQCD are sensitive to various
parameters, such as the parameters in wave functions of $B$ and
$D_{(s)}^{(*)}$. Because the wave functions are from
non-perturbative effect, we can not define them exactly. In
Table~\ref{tb:sensitivity}, we show examples of the sensitivity of
the branching ratios to parameters in (\ref{waved}). The
predictions of PQCD depend heavily  on  $\omega_D$ and  $a_D$,
which characterize the shape of $D_{(s)}^{(*)}$ wave function.

 \begin{table}[htbp]
\caption{The sensitivity of the decay branching ratios to change
of $a_D$($a_{D_s}$).} \label{tb:sensitivity}
\begin{center}
\begin{tabular}[t]{r|c|c}
 \hline
 \hline
 $a_{D_s}$ & $\mathrm{Br}(B^0 \to D_s^+ D_s^-)(10^{-5})$ \\
 \hline
 $0.2$ & $6.2$  \\
 $0.3$ & $7.8$  \\
 $0.4$ & $9.8$  \\
 \hline
 \hline
 $a_D$ & $\mathrm{Br}(B_s^0 \to D^+ D^-)(10^{-3})$ \\
 \hline
 $0.6$ & $3.3$ \\
 $0.7$ & $3.9$  \\
 $0.8$ & $4.5$  \\
 \hline
\end{tabular}
\end{center}
\end{table}

 From above discussion, the branching ratios within the reasonable
range of parameters in wave functions  are given as
\begin{equation}\begin{array}{ll}
\mathrm{Br}(B^0 \to D_s^+ D_s^-) = (7.8\pm_{1.6}^{2.0})\times
10^{-5},\ &
 \mathrm{Br}(B_s^0 \to D^+ D^-) =(3.6\pm 0.6 )\times 10^{-3};\\
\mathrm{Br}(B^0 \to D_s^{*+} D_s^-) =(6.0\pm_{1.1}^{1.6}) \times
10^{-5},\ &
 \mathrm{Br}(B_s^0 \to D^{*+} D^-) =(3.6\pm 0.6 )\times 10^{-3};\\
\mathrm{Br}(B^0 \to D_s^{*+} D_s^{*-}) = (8.5\pm_{1.8}^{2.0})
\times 10^{-5},\  & \mathrm{Br}(B_s^0 \to D^{*+} D^{*-})
=(6.4\pm1.2)\times 10^{-3}.
 \end{array}\end{equation}
Because $f_{D_s^*}$ is smaller than $f_{D_s}$, we can see the
branch ratio of $B^0 \to D_s^{*+} D_s^-$ is a little smaller than
$B^0 \to D_s^+ D_s^-$. The $\mathrm{Br}(B^0 \to D_s^{*+}
D_s^{*-})$ is larger than the others because of the extra
contribution from transverse polarization.

 In the decay $B\to \pi\pi$,
$m_\pi$ is much lighter than $m_B$. The energy release in the
decay is very large.  The final $\pi$ mesons runs very fast and
they may not have enough time to exchange the soft gluons and
resonance. Recently, the   $B\to D\pi$, $B\to D_s K(\phi)$ decays
with one heavy $D_{(s)}^{(*)}$ meson in final state are calculated
in PQCD approach \cite{lu:bdsk,dpi}. And the results are
consistent with the experiments, which shows that the final state
interaction may  not be important in those decays, although the
energy release is smaller than that in $B\to \pi\pi$ decays. Here
we also calculate $B^0 \to D_s^{(*)+} D_s^{(*)-}$ and $B_s^0 \to
D^{(*)+} D^{(*)-}$ decays with two heavy mesons in final states in
PQCD. We get large branching ratios comparable to other
predictions \cite{eeg}. This may be a hint for PQCD to work good
for these decays. The soft final state interaction in those
decays, for example, $B^0 \to D^+D^-$, and $D^+D^- \to D_s^+D_s^-
$ through exchanging $K^0(\bar{K^0})$ is somehow smaller than the
perturbative picture.

\begin{figure}[htbp]
\vspace{-0.5cm}
\begin{center}
\includegraphics[scale=1.0]{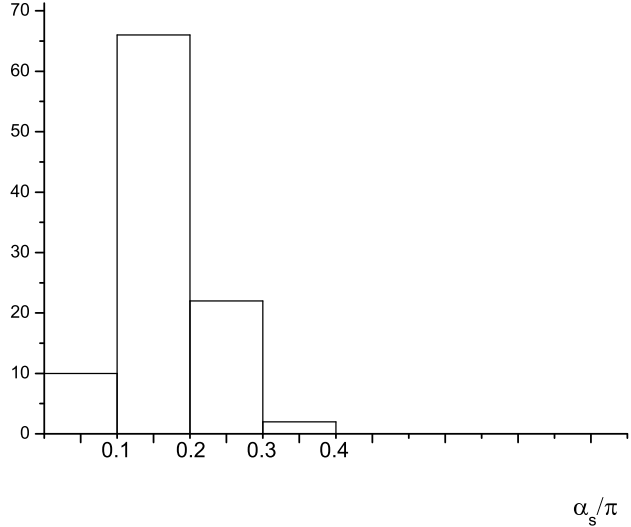}
\caption{Contributions to the branching ratio of decay $B^0 \to
D_s^+ D_s^-$ from different ranges of $\alpha_s/\pi$.}
\label{fig:diagrams3}
\end{center}
\end{figure}

For consistent check, in Figure \ref{fig:diagrams3}, we show the
contribution to the branching ratio of decay $B^0 \to D_s^+ D_s^-$
 from different ranges of $\alpha_s/\pi$, where the hard scale $t$
is given in appendix. From this figure, we find that most of
contribution comes from the range $\alpha_s/\pi<0.3$, implying
that the average scale is around $\sqrt{\Lambda_{QCD}m_B}$. It is
then numerically confirmed that PQCD may be even applicable to
$B^0 \to D_s^{(*)+} D_s^{(*)-}$ and $B_s^0 \to D^{(*)+} D^{(*)-}$
decays.

In ref.\cite{eeg}, J.O. Eeg et al.  computed those decays using
heavy-light Chiral quark model and their results read:
\begin{equation}\begin{array}{ll}
\mathrm{Br}(B^0 \to D_s^+ D_s^-) = 7.0\times 10^{-5},\ &
 \mathrm{Br}(B_s^0 \to D^+ D^-) =1.0\times 10^{-3}.
 \end{array}\end{equation}
Obviously, for decay $B^0 \to D_s^+ D_s^-$, we have the same
result. But we got different result for decay $B_s^0 \to D^+ D^-$
though our results are at the same order.  Unfortunately, there is
no direct experimental result about these decays up to now. We
hope those branching ratios will be measured soon in future and
these two theories can be tested.

\section{Conclusion} \label{sc:concl}

In this paper, we try to estimate the branching ratios of $B^0 \to
D_s^{(*)+} D_s^{(*)-}$ and $B_s^0 \to D^{(*)+} D^{(*)-}$ decays in
the heavy quark limit using perturbative QCD approach. These
decays can occur only through annihilation diagrams because the
four quarks in the final states are not the same as the ones in
$B$ meson. Our numerical results agree with the heavy-light chiral
quark model for  $B^0 \to D_s^{+} D_s^{-}$ and $B_s^0 \to D^{+}
D^{-}$ decays, which is not very small. We also give large
branching ratios for channels with one or two vector mesons in
final states. There is a hint for not large  soft final state
interactions.  We hope new experimental results will give a test
for  our results.

\section*{Acknowledgments}

 This work
was supported by National Science Foundation of China under Grants
No.~90103013, 10135060, 10475085, 10075013 and 10275035.


\end{document}